# Grete Hermann: An early contributor to quantum theory

## C. L. Herzenberg

**Abstract**:
The life and accomplishments of Grete Hermann are described. During the early twentieth century, she worked in physics, mathematics, philosophy and education. Her most notable accomplishments in physics were in the interpretation of quantum theory.

## Introduction

When we look back at the history of quantum mechanics, we find that some admirably accomplished individuals were involved who have since been largely forgotten. Among them was an interesting and very gifted woman, Grete Hermann, who worked in physics as well as mathematics, philosophy and education. In her work in mathematics and physics she was ahead of her time, and she carried out pioneering work with respect to the interpretation of quantum theory.

## Her life, briefly

Grete Hermann (also referred to variously as Margarethe Hermann, Margarethe Grete Hermann, Grete Henry-Hermann, Margaret Henry, Meg Henry Hermann, and Grete Henry) was actively involved in research and teaching in mathematics, physics, philosophy and education primarily during the early twentieth century. She was also politically active as a socialist, and, apart from the World War II period, lived much of her life in Germany.

Grete Hermann was born in Bremen, Germany, on March 2 1901. She was the third of seven children and grew up in a middle-class Protestant family (Fischer, Venz). Both of her grandfathers were pastors and her father was a merchant who seems to have become a street preacher in his later years; her mother, unlike her daughter Grete, was an intensely religious woman (Venz). Grete Hermann's early schooling was at the New School in Bremen (Venz).

Grete Hermann acquired teaching qualifications for secondary schools in 1921 and taught later at the Walkenmühle School (Fischer, Venz). Later in life she had a professorship at the Pedagogical Institute in Bremen.

From 1921 to 1925 she studied mathematics and philosophy at the universities in Göttingen and Freiburg. She was a doctoral student of Emmy Noether at Göttingen, and was awarded a Ph.D. in mathematics in 1926 (Kimberling, Jammer). Hermann followed her mentor Noether in working in both in mathematics and in physics.

Grete Hermann was also intensely concerned with philosophical issues, and worked as a private assistant to philosopher Leonard Nelson during 1926 and 1927 (Seevinck (a)). After Nelson's death in 1927, she continued work with his group, in particular with Minna Spect. Hermann published work on a system of philosophy of ethics and education in 1932.

She worked in Leipzig with Werner Heisenberg, Carl Friedrich von Weizsacker, and other prominent physicists during the early 1930s.

She is best known for discovering, in 1935, a logical error in John von Neumann's purported proof of the impossibility of hidden variables in quantum mechanics (Ray-Murray (a), (b); Wikipedia (a)). Although Hermann established that von Neumann's proof was invalid, her refutation of von Neumann's supposed proof was largely ignored for over 30 years.

In 1936 she was awarded the Richard Avenarius Prize by the Academy of Sciences of Saxony in Leipzig for her work on the significance of quantum theory and field theory of modern physics for the theory of knowledge (Fischer, Venz).

When she was a university student in Göttingen, Grete Hermann became active as a socialist. The philosopher Leonard Nelson with whom she worked had a political philosophy based on liberal socialism and took positions against social injustice and against the glorification of capitalism. He was involved in the founding in 1926 of a radical socialist youth group that she joined, the Internationalen Sozialistischen Kampfbund (ISK). Leading members of ISK included Willi Eichler, who was private secretary of Leonard Nelson and an organizer of ISK, and Minna Specht. The ISK received support from prominent European intellectuals, including Albert Einstein and Kathy Kollwitz. After Nelson's death in 1927, Hermann continued to collaborate with Minna Specht (Fischer). Starting in 1932, she worked as an editor for an associated daily newspaper "Der Funke" (Seevinck (a)) The ISK was involved in the formation of a united front against the Nazis; and subsequently was among the most active groups in the resistance during World War II (Venz). Grete Hermann also had a political commitment to the Internationalen Jugend Bund (International Youth Federation) (IJB), which had also been founded by Leonard Nelson (Venz).

An academic career was not possible for Grete Hermann under the conditions of the Nazi regime in Germany. However, she held classes for members of the resistance; topics including philosophy, politics and ethical values, and the ethics of the resistance against the Nazi regime were taken up and discussed (Fischer). Together with Leonard Nelson and Minna Specht, she helped to introduce a new, nonauthoritarian form of education in Germany (Venz). Leonard Nelson and Minna Specht and Grete Hermann were involved in opening a new school at Walkenmühle near Kassel, which was involved in both adult and childhood education, and included a socialist living community (Venz). The new educational efforts of Nelson, Specht and Hermann were brought to an end after Hitler's seizure of power, and the Walkenmühle School was closed by the Nazis in 1937 (Venz).

Persecution of the ISK group during 1935/6 caused some members of the ISK group to leave Germany. This persecution threatened Grete Hermann, and, like many others, she fled into exile.

By 1936, Hermann had left Germany for Denmark, where her friend Minna Specht had opened a school for children similar to the Wälkenmühle School that had been closed by the Nazis (Fischer, Venz, Wikipedia (a)). This socialist exile school ended up located in Østrupgaard, where Hermann lived (Mehra, Vol. 6, Part 2; Venz). Subsequently, the threat of war and the associated risk of a German occupation of Denmark threatened the survival of the exile school in Denmark. As a result, Grete Hermann left Denmark, and after a stay in Paris in 1938, she later left for exile in England in 1938, and went to live London (Fischer, Venz).

In 1937 she married Eduard Henry; this seems to have been largely a marriage of convenience to permit her to acquire British citizenship (Venz). During World War II she was involved with the underground resistance against the Nazi regime (Wikipedia). She lived in exile in England, and continued her activities in contact with Minna Specht and other colleagues. In exile, she was a leading member of the London ISK group; she worked at the ISK organization Sozialistische Warte. She dedicated herself completely to political work, and was involved in the programmatic discussion about the democratic rebuilding of Germany (Fischer, Venz). In 1941 she became a deputy member of the ISK on the executive committee of the Union Deutscher Sozialistischer Organisationen in Großbritannien (UNION) that brought together several socialist groups (Fischer, Venz). From April to August 1943 she was a member of the commission to draw up an action program to build a socialist united front.

In 1946, after World War II, she returned to Germany from England, and was subsequently divorced by Eduard Henry (Venz). In post-war Germany, she resumed the educational work that she had been involved in so many years earlier. She worked on the rebuilding and development of the Pädagogischen Hochschule in Bremen, and took over its leadership in1947. From 1950 to 1966, after her resignation as head of the Pädagogischen Hochschule, she had a professorship and taught on the faculty in philosophy and physics (Venz). She founded the Gewerkschaft für Erziehung und Wissenschaft (GEW), a trade union in education and science, and she was involved in educational and cultural activities with the Sozialdemokratischen Partei Deutschlands (SPD), the Social Democratic Party of Germany. From 1954 to 1966, she was a member of the Deutschen Ausschuss für das Erziehungs- und Bildungswesen (the German Committee for Upbringing and Education) (Fischer).

She continued her work in philosophy and wrote "Die Űberwindung des Zufalls – kritische Betrachtung zu Leonard Nelsons Begründung der Ethik als Wissenschaft".

From 1961 to 1978, she chaired the Philosophisch-Politischen Akademie (Philosophical-Political Academy) in Frankfurt, which had originally been initiated by the Leonard Nelson foundation and was an SPD-related institution (Fischer, Venz). Her philosophical commitment led to her becoming president of the Philosophical and Political Academy in Frankfurt (Venz).

During her later years, her interests reverted to politics and philosophy rather than mathematics and science. (After the death of her mentor Emmy Noether in 1935, and her being forced into exile in 1936, her creative work in science and mathematics was limited.)

Grete Hermann died on April 15, 1984, in her home city of Bremen, in Germany (Fischer, Venz).

**Her studies with Emmy Noether and her work in mathematics**

Grete Hermann studied mathematics in Göttingen, which was one of the main centers of mathematics in the world at that time; many distinguished mathematicians including Felix Klein, David Hilbert, and Emmy Noether taught in Göttingen then.

Grete Hermann became a protégé of mathematician Emmy Noether, who is recognized as one of the iconic figures of twentieth-century mathematics. The main line of Noether's mathematical research was the development of modern abstract algebra, and she was a key figure in the twentieth's century's trend toward abstraction in mathematics (Ars). Noether's work included a proof of the Lasker-Noether theorem that every ideal has a primary decomposition. The theorem had originally been proved by Emanuel Lasker using a difficult computational argument, but Noether identified the key abstract condition behind the result and used it to give a shorter proof of a much more general theorem. Rings that satisfy the ascending chain condition on ideals, the condition that her proof was based on, became known as Noetherian rings in her honor. Beyond her work in mathematics, Noether's insights into theoretical physics revealed the very important general connection between symmetries and conservation laws in physics (Byers, p. 67).

Hermann completed her doctorate under Noether (Kimberling; Jammer). While Noether was looking toward abstraction, the thesis that Hermann completed was rather a return to the computational approach that had been more widely used during the nineteenth century (Ars). Hermann showed that the approach used earlier by Emanuel Lasker could be turned into an effective procedure for computing primary decompositions (Ars). In her thesis, she demonstrated that Noether's proof of the Lasker-Noether theorem could be turned into an efficient algorithm for computing primary decompositions of polynomial ideals in Noetherian rings (Mravinci). Grete Hermann completed this work before the availability of computers, and even before the idea of an effective procedure had been formalized (Ars).

Grete Hermann's dissertation and 1926 paper "Die Frage der endlich vielen Schritte in der Theorie der Polynomideale" ("The question of finitely many steps in polynomial ideal theory") has been described as the foundational paper for computer algebra, and an intriguing example of ideas before their time (ACM; Wikipedia (a)). While computational aspects of mathematics had been somewhat more prevalent before the abstractions of the twentieth century took hold, mathematicians of that time appear to have been largely unfamiliar not only with the concept of computers but even of our contemporary concept of what an algorithm is (ACM). Her paper presents the first examples of procedures, with upper bounds given, for a variety of computations in multivariate polynomial ideals. Thus, Grete Hermann published a paper that to some extent anticipated by some 39 years the birth of computer algebra, which had been identified in the computing community as marked by a methodology invented and first presented in 1965 (ACM). Her thesis first established the existence of algorithms (including complexity bounds) for many of the basic problems of abstract algebra (Wikipedia).

Grete Hermann's was the first doctoral dissertation by a Noether student in Göttingen; she took her examinations in February 1925 with Noether and E. Landau (Dick, p. 51; MGP; Kimberling, p. 18-19, p. 40), and at age 24 was awarded a doctoral degree in mathematics from the Georg-August-Universität Gottingen (MGP; Venz). In later years, Grete Hermann remembered Emmy Noether, her "dissertation-mother," with great reverence, and delighted in cheerful memories of Noether (Dick, p. 51). Hermann followed her mentor Noether in working and developing significant results in both mathematics and physics. She also went on to work in philosophy and education.

**Her work in philosophy**

While in Göttingen, Grete Hermann worked as a private assistant to the philosopher Leonard Nelson from 1925 until his death in 1927, and she was an active member of the circle around Nelson (Jammer, p. 207). As a private instructor, Nelson had a position with the Department of Philosophy; his specialties included epistemology, ethics, education, legal doctrine, and policy. Nelson's approach to philosophy was based on those of the eighteenth century German philosopher Immanuel Kant and early nineteenth century philosopher and naturalist Jakob Friedrich Fries. Nelson was the founder of the Neo-Friesian school, and one of his goals was the development of Immanuel Kant's critical philosophy in the tradition of Jakob Fries (Heisenberg).

One of the requirements of the Friesian school and also of Nelson's circle was that all philosophical questions should be treated with the rigor of modern mathematics (Heisenberg). In the tradition of Kant, Nelson wanted to design a system of scientific ethics built up on a scaffolding of logical axioms. Nelson was a friend of the prominent mathematician David Hilbert, and in Hilbert's rigorous axiomatic justification of mathematics Nelson saw a model for building a systematic scientific ethics (Venz).

As a result of her association with Leonard Nelson, Grete Hermann was well informed in regard to the neo-Kantian ideas associated with Fries. As a neo-Kantian philosopher herself, Grete Hermann was conversant with transcendental idealism, as incorporated in the epistemology of Immanuel Kant and discussed in *The Critique of Pure Reason*. Kant presented it as a point of view which holds that our experience of things is about how they appear to us, not about those things as they are in and of themselves (Heisenberg; Wikipedia (c)).

Kant regarded knowledge as being structured out of sensory data and from universal concepts that he referred to as categories that the self imposes on the sensory data. Kant did treat ideas as being due to realities existing independent of human minds, but he regarded such independent realities or things-in-themselves as necessarily remaining unknown to us. He seems to have believed that human knowledge could not reach to such independent realities because knowledge could only be created in the course of synthesizing the ideas associated from sensory data.

Transcendental idealism can be clarified by examining Kant's account of how we intuit objects. Space and time, rather than being real things-in-themselves or empirically mediated appearances, are regarded as the forms of intuition by which we must perceive objects. They are therefore neither properties that we may attribute to objects in perceiving them, nor substantial entities in themselves. Instead, they are treated rather as subjective yet necessary preconditions of any object, to the extent that we regard that object is an appearance and not a thing-in-itself; as a result, we must necessarily perceive objects spatially and temporally (Wikipedia (c)).

This philosophical viewpoint evidently affected Grete Hermann's work in the interpretation of quantum mechanics. In view of Niels Bohr's concept of a phenomenon as irreducibly relative to a given experimental context, Hermann pointed out that, far from falsifying transcendental philosophy, quantum physics might be an incentive to radicalize it (Bitbol, p. 3).

As noted earlier, Hermann was greatly influenced by the philosopher Leonard Nelson, and subsequently she became the principal editor of Leonard Nelson's *Gesammelte Schriften* and thus was one of the principal individuals responsible for the perpetuation of Nelson's memory and work (Ross). This editing was not completed until late in her life, in the 1970s. This led to a fundamental break with the principles of Friesian epistemology which had been of concern to her earlier during her life (Ross).

**Her work in physics**

General

Grete Hermann had studied and completed her doctoral studies in mathematics at the Georg-August-Universität Göttingen, which had a distinguished faculty in physics and many eminent physicists as visitors, including Max Born, Werner Heisenberg, Enrico

Fermi, Pascual Jordan and other well-known physicists. During these years, a radical change in thinking in science was underway, a revolution in which classical Newtonian physics was being supplanted by quantum mechanics in much of the interpretation of the physical world.

Attempts to interpret and understand what was then new and very puzzling physics were of great contemporary importance when Grete Hermann entered the field, and her work in physics was mainly related to the interpretation of quantum mechanics. More specifically, her main work in physics was on the philosophical foundations of quantum mechanics, the significance of modern physics for the theory of knowledge, and causality in physics. She was one of the active early contributors to the historic debates on causality in quantum mechanics, and on the completeness of quantum mechanics and its description of reality (Mehra, Vol. 6, 1&2).

In the spring of 1934 she joined Werner Heisenberg's seminar as a visitor to the Physics Institute in Leipzig (Jammer, p. 207; AIP; Venz). In the early 1930s, Leipzig was one of the foremost centers for the study of quantum mechanics, and many notable scientists (including physicists Felix Bloch, Lev Landau, Rudolf Peierls, Edward Teller, Friedrich Hund, Carl Friedrich von Weizsacker, and also mathematician B. L. van der Waerden ) were present in Leipzig during that time period (Jammer, p. 207). Leipzig became well-known for activities in the study of the philosophical foundations and epistemological implications of quantum theory.

Heisenberg has written that Grete Hermann came to Leipzig for the express purpose of challenging the philosophical basis of atomic physics (Heisenberg). Heisenberg devoted an entire chapter of his book *Physics and Beyond: Encounters and Conversations* to a reconstruction of discussions that he had on quantum mechanics and Kantian philosophy with Grete Hermann and Carl Friedrich von Weizsäcker, and he indicated that both Hermann and von Weizsäcker contributed important insights to this topic (Heisenberg, Venz).

Grete Hermann seems especially to have worked together with C. F. von Weizsäcker and B. L. van der Waerden, and participated very actively in the Heisenberg seminar. As a result of this work in association with the Heisenberg seminar, in March 1935 Hermann was able to publish a lengthy and significant essay on the philosophical foundations of quantum mechanics, *Die naturphilosophischen Grundlagen der Quantenmechanik,* which contained her principal contributions to quantum mechanics and her major work in physics.

Her major contributions to physics were in examining the nature of causality in quantum mechanics, and in disproving a widely accepted but erroneous theorem (introduced by John von Neumann) that purported to prove the impossibility of hidden variable theory in quantum mechanics.

Carl Friedrich von Weizsäcker referred to her research the "first positive and undeniable contribution to the elucidation of the epistemological implications of quantum mechanics" (Fischer).

**Causality**

Grete Hermann had developed a significant interest in physics both from studying with Emmy Noether, whose work in mathematics had a highly significant impact on theoretical physics, and from working with the philosopher Leonard Nelson. She had developed a particular interest in issues relating to causality, and she became one of the active early contributors to the historic debates on causality in quantum mechanics (Mehra, Vol 6, 1&2).

In Grete Hermann's work, she attempted to maintain causality as a transcendental requirement for knowledge and experience, including in quantum mechanics. Her argument appears to have been based in part on Kant's Critique of Pure Reason.

Her point of departure in developing her major work was the empirical fact of the unpredictability of precise results in the measurement of microphysical objects (Jammer, p. 208).

Quantum mechanics, on which much of modern physics is based, makes only statistical predictions of the outcome of experiments. This is in contrast to classical Newtonian physics, which in principle at least makes definite predictions about where an object will be at any time in the future, given sufficiently accurate information about all of the forces acting on it and its initial conditions of position and momentum or velocity.

The new quantum physics, which generally only assigned probabilities to each of a range of possible outcomes of an experiment, presented a challenge to classical ideas of objectivity and causality, and was sometimes seen as forcing a renunciation of these fundamental classical concepts. Being aware that quantum mechanics seemed to imply a breakdown of universal causality, Grete Hermann approached the question by going to Leipzig and joining Heisenberg's seminar there, in the hope that she would be able to find a solution to this contradiction (Jammer, p. 207).

Her work in Leipzig led her to the relational interpretation of quantum mechanics (Jammer, p. 208). Bohr arrived considerably later at the relational concept of quantum mechanical states from considering the Einstein-Podolsky-Rosen thought experiment, and he became known for this approach, while Grete Hermann's work in introducing it has been all but forgotten (Jammer, p. 208).) In recent years, Rovelli has developed this approach more fully (Rovelli).

The term relational quantum mechanics refers to an interpretation of quantum theory which eliminates the concepts of the absolute state of a system, the absolute value of its physical quantities, or an absolute event. Relational theory describes only the way that systems affect each other in the course of physical interactions, and state and physical quantities always refer to the interaction, or the relation, between systems. However, the theory is regarded as being complete. In this approach, the physical content of quantum theory is be understood as an expression of the entire set of relations connecting all different physical systems (SEP).

As noted earlier, Hermann's point of departure in examining the philosophical foundations of quantum mechanics was from the empirical fact of the unpredictability of precise results in measurements on microphysical objects (Jammer, p. 208). The most apparent way out of such a situation would be by searching for a refinement of the quantum state description in terms of additional parameters ("hidden variables"). But that approach had appeared to be denied by quantum theory, at least according to John von Neumann's (erroneous) analysis of quantum theory (Jammer, p. 208). However, as discussed below, Hermann established the presence of a serious error in von Neumann's supposed proof of the impossibility of hidden variables in quantum mechanics. Since Hermann rejected von Neumann's supposed proof, she raised the question of what could justify the denial of additional parameters to the theory. Just because an adequate set of such parameters had not yet been developed did not appear to be adequate justification, as it might be said to violate the 'principle of the incompleteness of experience' (Jammer, p. 208). Hermann stated that the only sufficient reason for renouncing as futile any search for the causes of an observed result would be that the causes are already known (Jammer, p. 208).

As Max Jammer has pointed out, the dilemma faced in quantum mechanics then was this: Either the theory provides the causes which determine uniquely the outcome of a measurement (in which case, why should we not be able to predict the outcome?) or the theory does not provide such causes (but then how can the possibility of discovering them in the future be categorically denied?). Hermann saw the solution of this dilemma in the relational (or as she called it, the 'relative') character of the quantum mechanical description (Jammer, p. 208).

In her analysis, Hermann renounced the classical principle of objectivity. She replaced the idea of objectivity with the concept of dependency on instrumental measurement together with the idea that the physical process leading to a result can be causally reconstructed from the factual result of a measurement of a physical process. By this approach, Hermann explained why the theory prevents predictability without excluding a subsequent or after-the-fact identification of the causes of the particular outcome. Hermann presented a detailed explanation of how this could be achieved by examining in considerable detail a particular case, that of the von Weizsacker-Heisenberg thought experiment, which was an earlier precursor of the better known Einstein-Podolsky-Rosen thought experiment (Jammer, p. 208).

Hermann excluded the possibility of additional parameters on the grounds that quantum mechanics, though predictively indeterministic, could be considered retrodictively a causal theory (Jammer, p. 209). This situation can be rephrased to say that any additional causes or parameters would overdetermine a process and thus lead to a contradiction, since, with the final result of a measurement in sight, the causal sequence that led to the observed result could be reconstructed (Jammer, p. 209). Hermann distinguished between causality and predictability and emphasized the fact that they are not identical; she stated that "The fact that quantum mechanics assumes and pursues a causal account also for unpredictable occurrences proves that an identification of these two concepts is based on a confusion" (Hermann, quoted in Jammer, p. 209; Lenzen). This allows for the possibility that physical processes may be strictly determined even though exact prediction is not possible (Lenzen).

Grete Hermann showed that causality was retained in the sense that after an interaction, causes could be assigned for a particular effect. Von Weizsäcker expressed Hermann's conclusion by the statement that the persistence of classical laws can be applied to assign causes of past events but not to future events (Lenzen, p. 282).

Heisenberg seems to have approved of Grete Hermann's resolution of this dilemma; she quoted him as saying to her "That's it, what we were trying so long to clarify!" (Hermann letter quoted in Jammer, p. 208). However, Hermann's claim of retrodictive causality has been criticized by several authors, including Jammer, Stauss, and Buchel (Jammer, p. 209). Jammer regards her analysis of the von Weiszäcker-Heisenberg thought experiment as allowing the observed result without requiring it (Jammer, p. 209).

Hermann's views seem to emphasize the asymmetry between explanation and prediction in quantum mechanics as opposed to their symmetry in classical physics. This analysis was subsequently extended by others, including Norwood Russell Hanson, who appears to have emphasized that after a quantum event has occurred, a complete explanation of its occurrence can be given within the total quantum theory, but that it is in principle impossible to predict in advance those features of the event that can be explained after the fact (Jammer, p. 209).

Von Weizsäcker has described Grete Hermann's contributions as including clarifying that the impossibility of making certain predictions is not based on the fact that a causal chain investigated turns out to be interrupted somewhere, but rather on the fact that the different causal chains cannot be organized to form a unified picture embracing all aspects of the process, so that it remains to the choice of the observer which of the different causal chains is realized (Mehra, Vol. 6, part 2, p. 713).

Here is Grete Hermann's phrasing of it: "The difficulties in which the partisans of causality are placed by the discoveries of quantum mechanics, seem…not to arise from the causality principle itself. They rather emerge from the tacit assumption connected with it that the physical cognition grasps natural phenomena adequately and independently of the observational connection. This assumption is expressed in the prerequisite that every causal connection between processes yields a calculable action due to the cause, even more, that the causal connection is identical with the possibility of such a calculation." She added: "Quantum mechanics forces us to dissolve this mixing of different principles of natural philosophy, to drop the assumption of the absolute character of the cognition of nature, and to use the causal principle independently of the latter. By no means has it disproved causal law, but it has clarified its status and freed it from other principles which must not be combined with it necessarily." (quoted in Mehra, Vol. 6, part 2, p. 713).

**Determinism and the completeness of quantum mechanics: Von Neumann's error - Disproof of von Neumann's "No hidden variables theorem"**

Grete Hermann is best known for discovering a conceptual error in John von Neumann's purported proof of the impossibility of hidden variable theory in quantum mechanics. She was the first person to discover the error, and did so in 1935, three years after von Neumann published his supposed proof. While her discovery of the logical flaw in his work invalidated the theorem and allowed for the possibility of additional parameters, the so-called hidden variables, in quantum mechanics, her work was largely ignored rather than being promptly recognized and accepted. An unfortunate consequence of the lack of attention to her work was that the possibility of hidden variables continued to be regarded as impossible for many years afterwards.

John von Neumann was among the most distinguished and accomplished mathematicians of his time; in addition, he made major contributions to several other fields including quantum physics, and, most notably, was a pioneer of the application of operator theory to quantum mechanics. Von Neumann's abstract treatment of quantum theory empowered him to confront foundational issues, including that of determinism vs. non-determinism. But he made an unfortunate error: von Neumann, who had engaged in fundamental analyses of the basis of quantum mechanics, developed and presented a mathematical theorem purporting to show that no theory which assigns definite states to particles could be consistent. Otherwise stated, von Neumann's theorem purported to show that it is impossible for a theory with hidden variables to produce the predictions of quantum mechanics, and thus incorrectly 'proved' that a hidden variable theory of quantum mechanics could not exist.

A hidden variable theory is one that does not take the statistical predictions of the quantum formalism as the ultimate truth about reality, and that instead attempts to explain the random behavior of quantum mechanical systems in terms of unobserved deterministic variables. Thus, a hidden variable theory is an interpretation of the mathematics of quantum physics that does not necessarily accept the traditional Copenhagen interpretation in viewing nature as inherently random and the process of wave-function collapse as not being subject to further analysis beyond what existing equations of quantum mechanics tell us. (The traditional Copenhagen interpretation holds that reality is fundamentally random, and that, for example, the uncertainties in the position and momentum of a particle are as real as the position and momentum themselves (Ray-Murray (b)).) A hidden variable theory thus would give a physical interpretation of quantum phenomena at a deeper or subquantum level, giving a physical interpretation of the very nature of quantum systems, and construction of such a theory could provide a more explicit (and possibly even more fundamental) theory of quantum systems.

Quantum mechanics presents a probabilistic account of physical phenomena rather than a detailed causal account as classical mechanics does. The question of whether quantum mechanics should be considered a complete account of physical phenomena has a long and complicated history, and whether the account of the phenomena provided by quantum mechanics should be regarded as an exhaustive description of the physical reality associated with those phenomena has been addressed by many of the most original thinkers in physics (Guintini, p. 173). Could the statistical character of quantum mechanics be bypassed and eliminated by the introduction of supplementary parameters providing for a more detailed theory? Do statements of probability arise from the incompleteness of our knowledge, and could they be addressed more deeply with an explanation using hidden parameters, so-called hidden variables? Hidden variables might be introduced in the hope that they would provide a local deterministic objective description that would resolve or eliminate at least some of the paradoxical aspects of quantum mechanics.

The issue here was whether the quantum mechanical description of states might be regarded as encoding everything there is to know about the physical properties of systems, that is, whether quantum mechanics is a complete theory or not (Giuntini, p. 176). The question arose whether this could be done in a manner consistent with the mathematical constraints of the quantum mechanical formalism (Giuntini, p. 173). John von Neumann believed that an explanation by such hidden parameters would be incompatible with certain fundamental postulates of quantum mechanics (Giuntini, p. 173). Von Neumann seems to have regarded the concept of causality as equivalent to the concept of determinism, and indeterminism as a violation of causality (Giuntini, p. 178).

As noted earlier, in his book *The Mathematical Foundations of Quantum Mechanics*, published in 1932, von Neumann presented what purported to be proof of a theorem according to which quantum mechanics could not possibly be derived by statistical approximation from a deterministic theory of the type used in classical mechanics; thus, the aim of restoring causality in the quantum domain had been declared impossible to achieve (Giuntini, p. 181). Because of von Neumann's eminence, it was accepted by almost everyone in the field that he had proved what he thought that he had proved.

Von Neumann's impossibility claim amounted to a refutation of additional mathematical structures for quantum mechanics. After his formulation of the no-hidden-variables theorem, it was welcomed by the scientific community as the ultimate word on causality and determinism in the physical world (Giuntini, p. 174). The theorem was interpreted to show the impossibility of a hidden variable reconstruction of quantum mechanics. Von Neumann's attempted proof of the theorem was, however, invalid; it contained a conceptual error, as was first demonstrated by Grete Hermann, who discovered and published the flaw in this proof three years after its publication. Thus, a theory of quantum mechanics involving hidden variables might in fact be possible, contrary to von Neumann's claim (Ray-Murray (a); Gribbin).

While Grete Hermann does not seem to have been a proponent of hidden variable theory (she even is quoted as saying "that a deterministic supplement of quantum mechanics fails because quantum mechanics already allows us to give completely the causes for the occurrence of a given result of measurement" (quoted in Mehra, Vol. 6, Part 2, p. 733)); however, she identified the error in von Neumann's no-hidden-variables "proof". Hermann charged von Neumann with having constructed a fallacious argument in which the validity of the conclusion was assumed in the premise of the argument. She showed that he did this by introducing into the formal presuppositions of his proof a statement that was logically equivalent to the assertion to be proved. In von Neumann's "proof" that hidden-variables theories supposedly cannot work, he used that fact that a particular property of a quantum system obeys the commutative rules *on average*, and applied this rule to individual components of the quantum system, an incorrect argument (Gribbin). The issue in question concerned the additivity of the expectation value; that is, whether the expectation value of a sum is equal to the sum of the expectation values for arbitrary ensembles (Jammer, p. 273). In Section 7 of *Die naturphilosophischen Grundlagen der Quantenmechanik.*, Grete Hermann addressed the circularity of von Neumann's proof, and points out his assumption that the expectation value of the sum of two quantum mechanical quantities would equal the sum of the expectation values, noting that the validity of the proof depended on the validity of that assumption. She concluded that the additivity rule was too restrictive, and precluded the non-existence of dispersion free states (Seevinck, (b)). Grete Hermann commented on the problematic status of the additivity rule in the light of the impossibility of simultaneous measurement of noncommuting observables (Seevinck, (b)). According to Jammer, it appears that Hermann's criticism can be summarized to state that von Neumann had proved that the theory of quantum mechanics cannot be embedded in a hidden variable extension of the theory, but only as far as quantum mechanical states are concerned (Jammer, p. 273).

So, the lack of generality deriving from an assumption of linearity for expectation functionals seems to have been an essential problem with von Neumann's supposed proof. The von Neumann impossibility "proof" depended on a postulate of additivity; he essentially assumed that since the results of measurements are commutative on average in the quantum formalism, that the results of individual measurements should also be commutative (Ray-Murray, (a)). Von Neumann was simply wrong. Von Neumann's argument did not prove impossibility, as it claims. The argument failed due to its reliance on a physically unreasonable assumption (Wikipedia (2)).

The correct conclusion of von Neumann's argument would seem not to be that it precludes hidden variables under all circumstances, but rather that the proof is valid only for a very limited class of hidden variables (Seevinck (b), p. 19). Grete Hermann had shown that in all other circumstances, von Neumann's proof was invalid and hidden variable theories might be possible.

**Why did the scientific community ignore her work and its implications for so long?**

It is interesting (and something of an embarrassment for institutionalized science) that, although the von Neumann "proof" was actually refuted and its logical error exposed by Grete Hermann in 1935 (Ray-Murray (a)), few people at the time or for years thereafter seem to have paid attention to her work on this. Although Grete Hermann correctly identified a decisive error in John von Neumann's pivotal "proof" of the impossibility of hidden variables in quantum mechanics and she established that his supposed proof was invalid, that "proof" continued to be widely cited. Such was the momentum of the Copenhagen interpretation and von Neumann's reputation that when Grete Hermann pointed out that the supposed proof contained a blatant and devastating fallacy, she was simply ignored, and the Copenhagen interpretation remained the almost unquestioned accepted interpretation for decades (Goebel).

Grete Hermann's refutation of von Neumann's "proof" continued to be almost completely ignored for nearly 30 years, and the scientific community continued to tacitly accept and believe in von Neumann's false 'proof' until the flaw in his argument that had originally been discovered by Grete Hermann was rediscovered around 1965/6 by John Bell, who independently showed again that von Neumann's argument was based on a false assumption (Wikipedia (b); Gribbin; Rosenblum and Kuttner). Subsequently, in 1974, John Bell took up Grete Hermann's case further (Ray-Murray (a), (b)). John Bell's work enabled a wider appreciation of the unwarrantedly restrictive character of von Neumann's theorem; the upshot of Bell's 1966 paper was that a consistent formulation of hidden variable theories is entirely possible (Giuntini, p. 182).

Somewhat earlier than John Bell, David Bohm independently realized that von Neumann's theorem could only be relevant for a limited class of hidden variable theories (Giuntini, p. 182). In 1952, Bohm accomplished the 'impossible" by producing a counterexample to the by then long-accepted no-hidden-variables "proof" of von Neumann. Bohm thus showed that quantum theory was not inconsistent with the existence of real particles each with an actual position and an actual momentum. Bohm created a logically sound interpretation of quantum mechanics that included hidden variables, a fully-working hidden variables interpretation of quantum mechanics. However, "hidden variables" had by then become such an untouchable subject that Bohm's work was largely ignored by the physics community, surprising as that may seem in retrospect.

It was Bohm's work that inspired John Bell to challenge the no-hidden-variables "proof" and rediscover the error in von Neumann's supposed proof that had already been discovered so many years earlier by Grete Hermann (Rosenblum & Kuttner). Even after John Bell rediscovered the substance of Grete Hermann's refutation of von Neumann's supposed proof, some people in the field continued producing impossibility "proofs" with closely related errors at least as late as 1978, some 12 years after their central fallacy was brought to the fore by Bell, 26 years after a convincing counter-example had been shown to exist by Bohm, and 43 years after Grete Hermann first disproved von Neumann's "proof"(Ray-Murray (a)).

How could this happen? Here we have a flagrant example of communal dismissal of a correct result and continued disregard of it by the scientific community for nearly two generations. It is clearly of importance to understand the process by which progress takes place in science by the incorporation of correct scientific results into the accepted body of science. So it would seem to be worthwhile to consider the reception of Grete Hermann's work, and to ask why her criticism of von Neumann's 'proof' was ignored at the time, and also why her critique of that proof is still not widely known. This issue is of wider concern that Hermann's case in particular, as it illustrates several types of issues that have contributed to restricting progress in the development of science, and unfortunately continue to do so.

So, why was Grete Hermann's refutation of von Neumann's supposed proof ignored by the scientific community for so many years? Why did her result have no impact until it was rediscovered by John Bell some thirty years later? John Bell's result that he obtained 30 years later is much better known that Grete Hermann's original work. The overwhelming authority of the relatively well-established and esteemed von Neumann in contrast to the little-known female mathematician probably played a major role; as Feyerabend has noted, "the mere name 'von Neumann' and the mere word 'proof' silenced the objectors" (Seevinck (a), (b); Beller). This outcome is still somewhat surprising in view of the fact that Heisenberg and von Weizäcker were not only aware of Grete Hermann's refutation of von Neumann's supposed proof; actually they to some extent championed her ideas, and Heisenberg wrote a favorable preface to her leading publication. Furthermore, Pauli also realized that von Neumann's proof was deficient (Mara). However, the orthodox narrative continued to be constructed, and Heisenberg and the other establishment physicists continued to use von Neumann's supposed proof as an argument for indeterminism, and Grete Hermann's work fell by the wayside and did not achieve appropriate recognition or acceptance with the scientific establishment as a whole. That she phrased her analyses in a philosophical context and even in the specific philosophical traditions of Immanuel Kant, Herbert Fries, and Leonard Nelson may have been off-putting to some in the physics community (Mehra, Vol. 6, Part 2, p. 713). We could rephrase the question by asking why von Neumann's erroneous 'proof' of the impossibility of hidden variables in quantum mechanics was accepted by the scientific community for nearly 40 years after it was shown to be wrong by Grete Hermann, and why was her criticism ignored at the time, and even still not widely known? Furthermore, why was Grete Hermann's groundbreaking work on the relational concept of quantum mechanics all but forgotten, while Bohr became known for his approach to it?

We may be able to contribute to the further examination of these issues by listing some factors that may have contributed to influencing other scientists to ignore Hermann's work:

1. John von Neumann and Niels Bohr were well-recognized, high-ranking, and influential members of the scientific establishment, highly esteemed in the scientific community, with strong personalities. Their conclusions were sometimes quoted as gospel. Scientists, like other human beings, are herd animals, and tend to accept authority. If von Neumann had been slightly less powerful in his field, Grete Hermann's identification of the error in his conclusion might have been accepted instead of ignored, and the possibility of hidden variable interpretations or a deterministic version of quantum theory might then have been accepted as possible, rather than being rejected for so very many years. If Bohr had been slightly less powerful, Grete Hermann's introduction of relational quantum mechanics might have been ascribed to her instead of being subsumed as one of the accomplishments of Bohr.

2. Grete Hermann was not only young and without influential connections, but also she was an outsider, an interloper who had come into physics from a background in mathematics and philosophy.

3. Grete Hermann was little known in the wider physics community, outside the small core group of physicists addressing the fundamental scientific and philosophical issues relating to quantum theory, while von Neumann and Bohr were very well-known physicists.

4. Grete Hermann was an aspiring young woman at a time when almost all mathematicians, physicists, philosophers and other academics were male, and sex discrimination was extensively present in the academic and scientific establishment in many forms. The fact that she was female would not have worked in her favor.

5. Grete Hermann was also an outsider and political dissenter as a member of the political left, while the establishment was much more conservative. She was unable to obtain an academic position, and was subsequently forced into exile from Germany until after World War II. (Curiously enough, although perhaps coincidentally, David Bohm, whose later development of a fully valid hidden variables theory of quantum mechanics was ignored for many years, was also a political maverick, and politically to the left. After being called to testify by the House Un-American Activities Committee during the McCarthy witch-hunting years, he was suspended from his faculty position by Princeton University, and was unable to get another academic job in the United States, and then left the United States for what amounted to exile in Brazil (Rosenblum & Kuttner).)

6. Although Grete Hermann later in her life became a professor of physics and philosophy and taught physics courses, she seems not to have continued research in physics after she was forced into exile during World War II, and as a consequence she may have remained less known to physicists then engaged in research.

7. Grete Hermann published her work in physics in books and journals that were not that widely read (perhaps not by choice).

8. No English translation of even von Neumann's work became available for over 20 years, and translations of Hermann's work and writing about her in English came even later; this at a time when English was becoming or had become the international language of science as German had been before it.

9. Von Neumann's authoritative and difficult-to-read work is more quoted than studied, so that many physicists accepted it on the basis of evaluations by others and did not themselves examine it for flaws; while those who did examine his work and found possible errors may well have assumed on the basis of his authority that it was they who had made the mistakes; alternatively, they may have been unwilling to attack authority for fear of possible repercussions such as possible loss of credibility themselves, loss of influence, or loss of position. Similarly, Niels Bohr's statements were often vague and clouded in obscurity, difficult to analyze, and similar conclusions may apply.

10. Because of the pervasive influence of von Neumann's 'proof', for many years, many popular and semi-popular accounts of quantum theory and even many textbooks continued to state or imply that hidden-variable theories were impossible, even long after Bell's second refutation of von Neumann's arguments, so that students and the public as a whole were conditioned to believe in the impossibility of hidden-variable theories (Gribbin). The mental inertia of the physics community in accepting the interpretation to which they were initially exposed may have played a role in the continued acceptance of what had become the conventional interpretation of quantum mechanics and the rejection of deterministic hidden variable theories.

11. Grete Hermann's work in physics was at the boundaries between physics and philosophy and mathematics, an area not frequented by many practicing physicists.

12. Neils Bohr won the political war about the interpretation of quantum mechanics, even though physicists as important as Einstein disagreed with his interpretation. In science as in other human affairs, history is written by the victorious parties, and the history of quantum mechanics had been to a significant extent been written by physicists with a preference for the Copenhagen interpretation of quantum mechanics (Krogh). One consequence of this was that the contributions of Grete Hermann, some of whose work provided support to allowing an alternative approach to the interpretation of quantum mechanics, were marginalized.

13. Furthermore, even Grete Hermann's work in mathematics was also to some extent ignored for somewhat different reasons; this seemed to have occurred because her work in mathematics did not use what was then becoming a newly popular abstract approach but instead involved detailed computation and algorithm development; although later its relevance to the development of computational procedures and algorithms for computing became more widely appreciated.

14. Annoying as it may be for us to recognize our all-too-human limitations, it appears that scientists can be about as gullible as other people in accepting ideas simply because 'everybody knows' that they are true, and rejecting ideas at variance with orthodox establishment thinking (Gribbin). We can do better.

This listing of some of the factors that may have had a role in leading to the scientific community's ignoring Grete Hermann's significant contributions to physics, and thus to these contributions being effectively lost to science until rediscovered two generations later, is in all likelihood incomplete. It would be helpful if others could identify additional factors, not only in Hermann's case, but in other cases of discoveries being ignored by the scientific establishment for prohibitively long periods of time, to the detriment of scientific progress.

It has been argued that many aspects of contemporary physics would be interpreted differently today if historically a hidden variable interpretation of quantum mechanics had been presented first; and that the Bohm deterministic interpretation of quantum mechanics then would have been accepted as the dominant interpretation, and the Copenhagen interpretation would have been dismissed (Gribbin; Nicolić). Clearly, the history of physics would have turned out quite differently if Grete Hermann's work had been accepted in a timely manner.

Perhaps those of us in the communities of science and the history of science can learn further about what the cultural conditions for the retention rather than the rejection of new knowledge are, from a more thorough examination of Grete Hermann's case.

## References and Bibliography

grete hermann qm rev rev.doc
19 December 2008 draft